\documentclass[12pt]{article}

\usepackage{amsmath,amssymb,graphicx} 
\usepackage{epsf}
\usepackage{pstricks}
\usepackage{cite}

\newcommand{\beq}{\begin{eqnarray}}
\newcommand{\eeq}{\end{eqnarray}}

\newcommand{\centeron}[2]{{\setbox0=\hbox{#1}\setbox1=\hbox{#2}\ifdim

\wd1>\wd0\kern.5\wd1\kern-.5\wd0\fi
\copy0

\kern-.5\wd0\kern-.5\wd1\copy1\ifdim\wd0>\wd1
                                       \kern.5\wd0\kern-.5\wd1\fi}}
\newcommand{\ltap}{\>\centeron{\raise.35ex\hbox{$<$}}
                               {\lower.65ex\hbox{$\sim$}}\>}
\newcommand{\gtap}{\>\centeron{\raise.35ex\hbox{$>$}}
                               {\lower.65ex\hbox{$\sim$}}\>}

\newcommand\ZZ{\hbox{\zfont Z\kern-.4emZ}}
\font\zfont = cmss10 

\begin{document}
\begin{titlepage}
\begin{flushright}
{\tt hep-ph/0608266}
\end{flushright}

\vskip.5cm
\begin{center}
{\huge \bf Toward a Systematic Holographic QCD:\\

\vspace{.2cm}
A Braneless Approach
}

\vskip.1cm
\end{center}
\vskip0.2cm

\begin{center}
{\bf
Csaba Cs\'aki and Matthew Reece}
\end{center}
\vskip 8pt

\begin{center}
{\it Institute for High Energy Phenomenology\\
Newman Laboratory of Elementary Particle Physics\\
Cornell University, Ithaca, NY 14853, USA } \\
\vspace*{0.3cm}
{\tt  csaki@lepp.cornell.edu, mreece@lepp.cornell.edu}
\end{center}

\vglue 0.3truecm

\begin{abstract}
\vskip 3pt
\noindent
    Recently a holographic model of hadrons
motivated by AdS/CFT has been proposed to  fit the low energy data
of mesons. We point out that the infrared physics can be developed
in a more systematic manner by exploiting backreaction of the
nonperturbative condensates. We show that these condensates can
naturally provide the IR cutoff corresponding to confinement, thus
removing some of the ambiguities from the original formulation of
the model. We also show how asymptotic freedom can be incorporated
into the theory, and the substantial effect it has on the glueball
spectrum and gluon condensate of the theory. A simple
reinterpretation of the holographic scale results in a
non-perturbative running for $\alpha_s$ which remains finite for all
energies. We also find the leading effects of adding the higher
condensate into the theory. The difficulties for such models to
reproduce the proper Regge physics lead us to speculate about
extensions of our model incorporating tachyon condensation.
\end{abstract}

\end{titlepage}

\newpage


\section{Introduction}
\label{sec:intro}
\setcounter{equation}{0}
\setcounter{footnote}{0}

 QCD is a perennially problematic theory. Despite its decades of
experimental support, the detailed low-energy physics remains beyond
our calculational reach. The lattice provides a technique for
answering nonperturbative questions, but to date there remain open
questions that have not been answered. For instance, the low-energy
scalar spectrum is a puzzle. There are a lot of experimentally
observed states, however their composition (glueball vs. quarkonium)
and their mixings are not well understood. The difficulty for any
theory trying to make progress in this direction is to understand
the interaction between the scalar states and the vacuum condensates
of QCD. In this paper we attempt to incorporate the effects of the
vacuum condensates into the holographic model of QCD as a first step
toward understanding the scalar sector in the context of these
models.

The SVZ sum rules \cite{SVZ} are a powerful theoretical tool for
relating theoretically solid facts about perturbative QCD with
experimental data in the low-energy region. The basic observation is
that the correlator for a current $J$
\begin{equation}
\Pi(q) = i \int {\rm d}^4 x~e^{i q x}
\left<0\left|T{J(x)J(0)}\right|0\right>
\end{equation}
may be expanded in a Wilson OPE that is valid up to some power
(where instanton  corrections begin to invalidate the local
expansion), $(1/Q^2)({\rm d} \Pi/{\rm d}Q^2) = \sum C_{2d} \left<
{\cal O}_{2d}\right>$, where the ${\cal O}_{2 d}$ are
gauge-invariant operators of dimension $2d$. In the deep Euclidean
domain the coefficients $C_{2d}$ are calculable. On the other hand,
the correlator relates to observable quantities; for instance if $J$
is the vector current then ${\rm Im} \Pi(s)$ is directly
proportional to the spectral density $\rho(s)$, measurable from the
total cross section $\sigma (e^{+} e^{-} \rightarrow {\rm
hadrons})$. The SVZ sum rules, then, relate Wilson OPE expansions to
measurable quantities. They can become useful for understanding
detailed properties of the first resonances when one takes a Borel
transform that suppresses the effect of higher resonances. As it
turns out, keeping only low orders of perturbation theory in the
coefficients $C_{2d}$, one still obtains reasonable agreement with
data, so the assumption that the largest corrections arise from the
condensates is a good one.

Recently another technique for understanding the properties of
low-lying  mesons has arisen in the form of AdS/QCD. The
phenomenological model constructed on this basis \cite{AdSQCD} takes
as its starting points the OPE (much as in the SVZ sum rules) and
the AdS/CFT correspondence \cite{AdSCFT}. The idea is
straightforward: rather than attempting to deform the usual Type IIB
on $AdS_5 \times S^5$ to obtain a theory more like QCD, one starts
with QCD and attempts to build a holographic dual. Of course in
detail such a program is bound to eventually run up against
difficulties from $\alpha^{\prime}$ corrections, $g_s$ corrections,
the geometry of the five compact dimensions (or the proper
definition of a noncritical string theory) and other issues.
However, one can set aside these problems, begin with a relatively
small set of fields needed to model the low-lying states in QCD, and
see how well the approximation works. In this phenomenological
approach with bulk fields placed in the Randall-Sundrum background
\cite{RS}, and the $AdS$ space cut off with a brane at a fixed $z =
z_c$ in the infrared a surprisingly good agreement with the physics
of the pions, $\rho$, and $a_1$ mesons has been found~\cite{AdSQCD}
(for more recent work on AdS/QCD
see~\cite{MoreAdSQCD,HirnRiusSanz}). However, there are several
obvious limitations to this model. For example, it does not take
into account the power corrections to the OPE in the UV (the effects
of the vacuum condensates), or the corrections coming from
logarithmic running. Also, the theory has a single mass scale (set
by the location of IR cutoff brane) which determines the mass scales
in all the different sectors in QCD.

Here we show how the effect of the vacuum condensates and of
asymptotic freedom can be simply incorporated into the model. We
will limit ourselves to pure Yang-Mills theory without quarks,
though we expect that most of the aspects of incorporating quarks
should be relatively straightforward. For the vacuum condensates one
has to introduce a dynamical scalar with appropriate mass term and
potential coupled to gravity. There will be a separate field for
every gauge invariant operator of QCD, and a non-zero condensate
will lead to a non-trivial profile of the scalar in the bulk. While
the effects of these condensates on the background close to the UV
boundary are small (though not always negligible), they will become
the dominant source in the IR and effectively shut off the space in
a singularity (and without having to cut the space off by hand).
This resolves several ambiguities in choosing the boundary
conditions at the IR brane, and the various condensates will also be
able to set different mass scales in the different sectors of QCD.
Asymptotic freedom can be achieved by properly choosing the
potential for the scalar corresponding to the Yang-Mills gauge
coupling. Incorporating asymptotic freedom will have an important
effect on the glueball sector since a massless Goldstone field will
pick up a mass from the anomalous breaking of scale invariance.

It is natural to ask why we should expect QCD to have any useful
holographic dual. The most well understood examples
of holography are in the limit of large $N$ and $g_{YM}^2 N \gg 1$,
far from the regime of real-world QCD, which apparently would
be a completely intractable string theory with a large value
of $\alpha'$. On the other hand, the large $N$ approximation
has frequently been applied in QCD phenomenology.
A major concern in applying holography
to QCD is whether the dual should be local, or whether it can
have higher-derivative $\alpha'$ corrections.
The $\alpha'$ corrections are associated with
the massive stringy excitations in the bulk, which once integrated
out yield complicated Lagrangians for the remaining light fields.
However, we can think of this physics in a different way. Holography
is closely related to the renormalization group~\cite{HoloRGE}; the
coordinate $z$ can be identified with $\mu^{-1}$, where $\mu$ is
the renormalization group scale. AdS/CFT identifies massive bulk
fields with higher-dimension operators in the field theory. From this
point of view, $\alpha'$ corrections to the physics of light bulk fields
are associated with the effects of higher-dimension operators coupled
to the low-dimension operators in the RG flow. The OPE tells us
that, at large $Q^2$, the effects of these higher-dimension operators
in the field theory are controllably small. From this point of view,
despite the apparently large $\alpha'$ corrections, it is reasonable
to begin with a local bulk action in terms of fields corresponding
to the low dimension operators of QCD. The example of SVZ
gives us hope that this can correctly capture physics of light mesons.
For highly excited mesons, which in QCD look like
extended flux tubes, and thus feel long distances, it is more
probable that the strong IR physics will mix different operators
and that our neglect of $\alpha'$-like corrections will become
more troublesome. It is a general problem, in fact, that physics of
highly excited hadrons is troublesome in these models~\cite{Shifman05}.
Recent proposals for backgrounds with correct Regge physics~\cite{KarchRegge}
offer some hope of addressing this problem. We will offer some
further comments on how a closed-string tachyon might give a
dynamical explanation of such a background, but there is clearly
much more to be done along these lines.

The paper is organized as follows. In section 2 we will remind the
reader of the basic formulation of AdS/QCD  on Randall-Sundrum
backgrounds as in Ref.\cite{AdSQCD}, and discuss some of its
shortcomings. In section 3 we will discuss a class of models that do
not incorporate the running of the QCD coupling, but do incorporate
the effects of the lowest vacuum condensate. We will point out that
these backgrounds too have some shortcomings, but improve on the RS
backgrounds, and may provide a useful setting for exploring some
questions. We calculate the gluon condensate and point out that
there is a zero mode in the glueball spectrum due to the spontaneous
breaking of scale invariance. In section 4 we discuss the
construction of 5D theories with asymptotic freedom. We calculate
the gluon condensate and provide a first estimate of masses for the
$0^{++}$ glueballs from these backgrounds. We point out a possible
interpretation of the running coupling where $\alpha_s$ remains
finite for all energies. In section 5 we show how to systematically
incorporate the effects of the higher condensates. We give a
background including the effect of $\langle {\rm Tr F^3} \rangle$
and show how it affects the gluon condensate and the glueball
spectrum. In section 6 we discuss the difficulties of reproducing
the correct Regge physics, and speculate that closed string tachyon
dynamics could perhaps be responsible for reproducing the necessary
IR background. Finally we provide the conclusions and some outlook
in section 7.

\section{AdS/QCD on Randall-Sundrum backgrounds}
\setcounter{equation}{0}
\setcounter{footnote}{0}

We briefly review the AdS/QCD model on Randall-Sundrum backgrounds
\cite{AdSQCD}.  One assumes that the metric is exactly AdS in a
finite region $z = 0$ to $z = z_c$, i.e.
\begin{equation}
ds^2 = \left(\frac{R}{z}\right)^2 \left(dx^\mu dx^\nu \eta_{\mu \nu}
- dz^2\right),~0 \leq z\leq z_c,
\end{equation}
 where $z_c \sim
\Lambda^{-1}_{QCD}$ determines the scale of the mass spectrum. It is
assumed that there is a brane (the ``infrared brane") at $z = z_c$;
in practice one assumes a UV boundary at $z = \epsilon$ and sends
$\epsilon \rightarrow 0$ at the end of calculations. One puts a
field $\phi_{\cal O}$ in the bulk for every gauge invariant operator
${\cal O}$ in the gauge theory. If ${\cal O}$ is a $p$-form of
dimension $\Delta$, $\phi$ has a 5D mass $m_5^2 = (\Delta +
p)(\Delta + p - 4)$. This operator has UV boundary conditions
dictated by the usual AdS/CFT correspondence. On the other hand, IR
boundary conditions are less clear, and one in principle can add
localized terms on the IR brane.

For instance, in the original papers the treatment involves the rho
and $a_1$ mesons and the pions. There are bulk gauge fields, $A^L_M$
and $A^R_M$ (where $M$ is a 5D Lorentz index) coupling to the
operators $\bar{q}_L \gamma^\mu t^a q_L$ and $\bar{q}_R \gamma^\mu
t^a q_R$. There is also a bulk scalar $X^{\alpha \beta}$ coupling to
the operator $\bar{q}^{\alpha}_R q^{\beta}_L$, where $\alpha$ and
$\beta$ are flavor indices. The scalar $X^{\alpha \beta}$ is {\em
assumed} to have a profile proportional to $\delta^{\alpha \beta}
\left(m_{\alpha} z + \left<q_{\alpha} \bar{q}_{\alpha}\right>
z^3\right)$, based on the Klebanov-Witten result \cite{KW} that a
nonnormalizable term in a scalar profile corresponds to a
perturbation of the Lagrangian by a relevant operator (in this case,
$m\bar{q}q$), while a normalizable term corresponds to a
spontaneously generated VEV for the corresponding field theory
operator. The profile for $X(z)$ couples differently to the vector
and axial vector mesons and achieves the $\rho - a_1$ mass
splitting. Furthermore, the pions arise as pseudo-Nambu-Goldstone
modes of broken chiral symmetry, and the Gell-Mann--Oakes--Renner
relationship is satisfied. Several other constants in the chiral
Lagrangian are determined to around 10\% accuracy.

There are several limitations to this model. One is that it does not
take into  account the power corrections to the OPE in the UV, or
the corrections coming from logarithmic running. The SVZ sum rules
work fairly well with just the leading power correction taken into
account~\cite{SVZ}, and these power corrections can be incorporated
into the form of the metric with a simple ansatz\cite{HirnRiusSanz}.
However, backreaction has not been taken into account in such
studies, so the Einstein equations will not be exactly satisfied.
Also, in the SVZ sum rule approach, leading corrections to the OPE
essentially determine the mass scale of the lightest resonance in
the corresponding channel. In the Randall-Sundrum approach, the
leading correction to the OPE and the IR wall at $z_c$ {\em both}
influence the corresponding mass scale. This has effects on the
spectrum.

For instance, in QCD the mass scales associated with mesons made
from  quarks and mesons made from gluons are very
different~\cite{HadronsAlike}. One can see this in the OPE. For
instance, in the tensor $2^{++}$ channel, the leading corrections to
the OPE come from the same operator in the $q\bar{q}$ case as in the
$G^2$ case, but the coefficients are very different. In the
Randall-Sundrum approach, without turning on a background field VEV
the mass scale for both of these channels is set by $z_c^{-1}$. One
can turn on a background VEV and give the fields different couplings
to it to reproduce the difference in the OPEs, but $z_c^{-1}$ will
continue to play a role in setting their masses.

Another limitation is that these backgrounds lack asymptotic
freedom. If one defines the theory with a cutoff at which $\alpha_s$
has some finite value, this might not appear to be of central
importance. On the other hand, we understand that in real QCD the
values of the condensates are determined by the QCD scale at which
the perturbative running coupling blows up, so achieving asymptotic
freedom can allow such a relationship and make the model less {\em
ad hoc}. Furthermore, we will see that the lack of proper conformal
symmetry breaking can lead to a massless scalar glueball state
(i.e., the model has a ``radion problem"), which is most
satisfactorily resolved by incorporating asymptotic freedom.

In summary, the AdS/QCD models on hard-wall backgrounds work
surprisingly  well for some quantities, but have obvious drawbacks.
There are ambiguities in IR boundary conditions, and the existence
of a single IR wall influencing all fields obscures the relationship
between masses of light resonances and power corrections discovered
by Shifman, Vainshtein, and Zakharov. Luckily, there is a simple
remedy to these difficulties: we simply remove the IR brane, and
allow the growth of the condensates to dynamically cut the space off
in the infrared.

\section{Vacuum condensates as IR cutoff}
\setcounter{equation}{0}
\setcounter{footnote}{0}

To model the pure gauge theory we begin with the action for
five-dimensional gravity coupled to a dilaton (in the Einstein
frame):
\begin{equation}
S = \frac{1}{2\kappa^2} \int {\rm d}^5 x \sqrt{g} (-{\cal R} +
\frac{12}{R^2}  +\frac{1}{2} g^{MN} \partial_{M} \phi \partial_{N}
\phi). \label{action}
\end{equation}
Here $\kappa^2$ is the 5 dimensional Newton constant and $R$ is the
AdS curvature (related to the the (negative) bulk cosmological
constant as $R^{-2}=-\frac{\kappa^2}{6}\Lambda$. Note, that $\phi$
is dimensionless here. The dilaton will couple to the gluon operator
$G_{\mu\nu}G^{\mu\nu}$. The fact that there is a non-vanishing gluon
condensate in QCD is expressed by the fact that the dilaton will
have a non-trivial background. We can find the most general such
background by solving the coupled system of the dilaton equation of
motion and the Einstein's equation, under the ansatz that we
preserve four-dimensional Lorentz invariance while the fifth
dimension has a warp factor:
\begin{eqnarray}
&& ds^2= e^{-2 A(y)} \eta_{\mu\nu}dx^\mu dx^\nu-dy^2 \nonumber \\
&& \phi =\phi (y).
\end{eqnarray}
The coupled equations for $A(y),\phi (y)$ will then be
\begin{eqnarray}
&& 4 A'^2-A''= 4 R^2 \nonumber \\
&& A'^2= \frac{\phi'^2}{24}+\frac{1}{R^2} \nonumber \\
&& \phi''=4 A' \phi'.
\end{eqnarray}
A simple way of solving these equations is to use the superpotential
method~\cite{Superpotential,CEGH},  that is define the function
$W(\phi )$ such that
\begin{eqnarray}
&& A'(y)= W (\phi (y)) \nonumber \\
&& \phi'(y)= 6 \frac{\partial W}{\partial \phi}.
\label{Weqs}
\end{eqnarray}
This is always possible, and the equation determining the superpotential is given by
\begin{equation}
V=18 \left( \frac{\partial W}{\partial \phi} \right)^2-12
W^2=-\frac{12}{R^2}. \label{pot}
\end{equation}
To solve for the most general superpotential consistent with our
chosen (constant)  potential, it is useful to parameterize it in
terms of a ``prepotential'' $w$ as~\cite{CEGH}
\begin{eqnarray}
&& W=\frac{1}{R}\left( w+\frac{1}{w}\right) \nonumber \\
&& W'=\sqrt{\frac{2}{3}} \frac{1}{R}\left( w-\frac{1}{w}\right),
\label{prepot}
\end{eqnarray}
which is chosen such that (\ref{pot}) is automatically satisfied.
The consistency  condition of the two equations in (\ref{prepot})
implies a simple equation for the prepotential:
\begin{equation}
w'=\sqrt{\frac{2}{3}} w,
\end{equation}
and thus for the superpotential we find
\begin{equation}
W(\phi )=\frac{1}{R}( C e^{\sqrt{\frac{2}{3}} \phi}+ C^{-1} e^{- \sqrt{\frac{2}{3}}\phi} ).
\end{equation}
With the superpotential uniquely determined (up to a constant $C$) we can then go ahead and
integrate the equations in (\ref{Weqs}). The result is
\begin{eqnarray} \label{eq:solution1}
&& \phi (y)=\sqrt{\frac{3}{2}} \log \left[ C \tanh 2 (y_0-y)/R \right], \nonumber \\
&& A(y)= -\frac{1}{4} \log \left[ \cosh 2 (y_0-y)/R \sinh 2 (y_0-y)/R \right]+A_0,
\end{eqnarray}
where $y_0$ and $A_0$ are integration constants. These are the
solutions also found in~\cite{CEGH,dilatonbackground}. In order to
have $A(y)$ asymptotically equal to $y$ (for large negative $y$), we
will fix $A_0 = \frac{y_0}{R}-\frac{1}{2}\log 2$. Note, that there
is another branch of the solution where in which $\tanh$ is replaced
by $\coth$ in the solution for $\phi$. In order to have a form of
the solution that is more familiar and useful, we make a coordinate
transformation $e^{\frac{y-y_0}{R}}=z/z_c$. This will recast the
solution in a form that matches the usual conformal coordinates
$x^{\mu}, z$ near $z = 0$:
\begin{eqnarray}
{\rm d}s^2 & = & \left(\frac{R}{z}\right)^2 \left(\sqrt{1 - \left(\frac{z}{z_c}\right)^8} \eta_{\mu \nu}
{\rm d}x^\mu {\rm d}x^\nu - {\rm d}z^2 \right) \\
\phi(z) & = & \sqrt{\frac{3}{2}} \log \left(\frac{1 + \left(\frac{z}{z_c}\right)^4}{1 -
\left(\frac{z}{z_c}\right)^4}\right)+\phi_0.
\label{solution}
\end{eqnarray}

Here $z_c$ is a new parameter determining the IR scale, and we expect $z_c \sim \Lambda_{QCD}^{-1}$.
The point $z = z_c$ is a naked
singularity, which we must imagine is resolved in the full string theory. Also note that the first correction to
the $AdS_5$ metric goes as $z^8$.

\subsection{Gluon condensate}

We have seen that the metric incorporates power corrections to the
pure  AdS solution, which we want to identify with the effects of
the gluon condensate. Below we would like to make this statement
more precise. According to the general rules of the AdS/CFT
correspondence given some field $\varphi_{\cal O}$ on this
background representing an operator ${\cal O}$ in QCD, the
deep-Euclidean correlator of ${\cal O}$ will have a $Q^{-8}$
correction if $\varphi_{\cal O}$ has no dilaton coupling, and a
$Q^{-4}$ correction if $\varphi_{\cal O}$ couples to $\phi$. Note
that near $z = 0$, the dilaton behaves $\phi_0+
\sqrt{6}\frac{z^4}{z_c^4}$. This is in agreement with the
expectation that a field coupling to an operator with dimension
$\Delta$ has two solutions, $z^{\Delta -d}$ and $z^\Delta$. In our
case $\Delta =d=4$, and we expect the constant piece to correspond
to the source for the operator ${\rm Tr}G^2$ and coefficient of the
$z^4$ to give the gluon condensate. The precise statement~\cite{KW}
from AdS/CFT is that if solution to the classical equations of
motion $\Phi$ has the form near the boundary
\begin{equation}
\Phi (x,z) \to z^{d-\Delta} [\Phi_0 (x)+{\cal O}(z^2)]+z^\Delta [A(x)+{\cal O}(z^2)]
\end{equation}
then the condensate (one-point function) of the operator ${\cal O}$ is given by
\begin{equation}
\langle {\cal O}(x) \rangle = (2 \Delta -d) A(x) .
\label{cond}
\end{equation}
However, to apply this to the solution in (\ref{solution}) we need
to make sure that the appropriate normalization of the fields is
used. The expression (\ref{cond})  is derived assuming a scalar
field action of the form $1/2 \int d^4xdz/z^3 (\partial_M \phi)^2$.
Comparing this with the action used here (\ref{action}) we find
\begin{equation}
\langle {\rm Tr} G^2 \rangle = 4 \sqrt{3} \sqrt{\frac{R^3}{\kappa^2}} \frac{1}{z_c^4}.
\end{equation}

In order to be able to relate this expression for the condensate we
need to find an  expression for $R^3/\kappa^2$. This can be done by
requiring that the leading term in the OPE of the gluon operator
$G^2$ is correctly reproduced in the holographic theory. One can
simply calculate the leading term in the OPE in QCD~\cite{Kataev}
\begin{equation}
\int d^4x \langle G^2 (x) G^2 (0) \rangle e^{i q x} = -\frac{(N^2-1)}{4 \pi^2} q^4 \log \frac{q^2}{\mu^2} +\ldots
\end{equation}
The same quantity can be calculated in the gravity theory by
evaluating the action for the  scalar field with a give source
$\phi_0 (q)$ in the UV. The general expression for the action is
obtained (after integrating by parts and using the bulk equation of
motion)
\begin{equation}
S_{5D}=\frac{1}{2\kappa^2} \int d^4 x \frac{R^3}{z^3} \frac{1}{2} \phi \partial_z \phi |_{z\to 0}.
\end{equation}
In order to find the action one can use the bulk equation of motion close to the UV for the scalar field
given by
\begin{equation}
\phi''+q^2 \phi -\frac{3}{z} \phi'=0 .
\end{equation}
Requiring that the wave function approaches 1 around $z=0$ will fix the leading terms in the wave function:
\begin{equation}
\phi (z) = 1-\frac{1}{32} q^4 z^4 \log q^2 z^2  +\frac{1}{4} q^2 z^2 +\ldots .
\end{equation}
Taking the second derivative of the action we find
\begin{equation}
\int d^4x \langle G^2 (x) G^2 (0) \rangle e^{i q x} = -\frac{R^3}{16\kappa^2} q^4 \log q^2/\mu^2 +\ldots ,
\end{equation}
from which we get the identification
\begin{equation}
\frac{R^3}{\kappa^2}= \frac{4 (N^2-1)}{\pi^2}.
\label{kap2}
\end{equation}
Using this result we get a prediction for the gluon condensate
\begin{equation}
\langle {\rm Tr} G^2 \rangle = \frac{8}{\pi z_c^4} \sqrt{3 (N^2-1)}.
\end{equation}

\subsection{The Glueball Spectrum}

Now we wish to solve for the scalar glueball spectrum, which is
associated with scalar fluctuations  about the dilaton--metric
background. The analogous calculation in the supergravity model has
been performed in~\cite{SugraBH}. We should solve the coupled
radion--dilaton equations, as often there is a light mode from the
radion. In other words, we should solve for eigenmodes of the
coupled Einstein-scalar system. This has been worked out in detail
for a generic scalar background in \cite{radion}: the linearized
metric and scalar ansatz is given by:
\begin{eqnarray}
&& ds^2 = e^{-2 A(y)}(1-2 F(x,y)) dx^\mu dx^\nu \eta_{\mu\nu} -(1+ 4
F(x,y)) dy^2 \nonumber \\
&& \phi (x,y)= \phi_0 (y) + \frac{3}{\kappa^2 \phi_0} \left(
F'(x,y)-2 A'(y) F(x,y) \right).
\end{eqnarray}
This will satisfy the coupled Einstein-scalar equations if $F= F(y)
e^{i q\cdot x}$ with $q^2=m^2$ and $F(y)$ satisfies the differential
equation
\begin{equation} \label{eq:radion}
F''-2A'F'-4A''F -2 \frac{\phi_0''}{\phi_0'}F'+4A'
\frac{\phi_0''}{\phi_0'} F=-e^{-2A} m^2 F .
\end{equation}
Using the solutions for $A(y)$ and $\phi_0(y)$ from
Eq.~\ref{eq:solution1}, and  an ansatz $F(x,y) = F(y) e^{i q \cdot
x}$, with $q^2 = - m^2$, this becomes:
\begin{equation}
F''(y) + \frac{10}{R} \coth\frac{4(y-y_0)}{R} F'(y) +
\left(\frac{16}{R^2} + m^2  \frac{e^{2 y_0/R}}{\sqrt{2 \sinh
\frac{4(y_0-y)}{R}}}\right) F(y) = 0.
\end{equation}
Demanding a normalizable solution, we need that (in the $z$
coordinates) $\int dz \sqrt{g}  |\varphi(z)|^2$ and $\int dz
\sqrt{g} g^{55} |\partial_z \varphi(z)|^2$ be finite. Thus we need
$\varphi(z) \sim z^4$ at small $z$. We need to be somewhat more
careful about solving for $F$: in the $z$ coordinates, we have that
\begin{equation}
z \rightarrow 0: z \frac{dF}{dz} - 2 z \frac{dA}{dz} F \rightarrow z
\frac{dF}{dz} - 2 F \sim \varphi(z), \label{BCglueball}
\end{equation}
so that $F(z) \sim z^2$ near $z = 0$ is compatible with our
assumptions on the behavior  of $\varphi(z)$. As usual, this
equation can be solved using the shooting method: the differential
equation is solved numerically starting from the UV boundary with
arbitrary normalization the BC following from (\ref{BCglueball})
(the numerics is very insensitive to the choice of the actual UV BC
for the higher modes) for varying values of $m^2$. For discrete
values of $m^2$ the wave function will be normalizable (which
numerically is equivalent to requiring a Neumann BC at the location
of the singularity). This way we find (in units of $z_c^{-1}$)
glueballs with masses $6.61$, $9.84$, $12.94$, and $15.98$ (and so
forth, with regular spacing in mass).

There is also a serious problem: we find a massless mode for the
radion (with the  $F(z) \sim z^2$ UV boundary conditions). This can
be understood as follows: in real QCD, the classical conformal
symmetry is broken by the scale anomaly. However, in our model, it
is broken by the $z^4$ profile of the dilaton. AdS/CFT tells us we
should understand turning on such a normalizable background in the
UV as a spontaneous symmetry breaking. In Randall-Sundrum models,
one similarly has a radion problem and needs to invoke (for
instance) a Goldberger-Wise stabilization \cite{GoldbergerWise} to
avoid a massless mode. The radion has not been part of previous
investigations of glueballs on Randall-Sundrum backgrounds
\cite{BoschiFilhoBraga}, so such studies have essentially assumed
that the stabilization mechanism removes a light mode from the
spectrum. However, an added Goldberger-Wise field does not seem to
correspond to an operator of QCD, so it goes against the spirit of
the AdS/CFT correspondence. It is apparent that a palatable solution
of the radion problem in AdS/QCD demands a 5D treatment of scale
dependence that mirrors that in 4D. This motivates us to search for
backgrounds incorporating asymptotic freedom, which also allows us
to begin approaching a more detailed matching to perturbative QCD.

\section{Incorporating Asymptotic freedom}
\label{sec:asymfree}
\setcounter{equation}{0}
\setcounter{footnote}{0}

We have shown in the previous section a background that incorporates
the lowest QCD condensate ${\rm Tr} F^2$ and which automatically
provides an IR cutoff via the backreaction of the metric. However,
this setup is certainly too simplistic even to just produce the main
features of QCD: for example asymptotic freedom is not reproduced in
that setup.  It is the dilaton field that also sets the QCD coupling
constant, and by approaching a constant value the model in the
previous section actually describes a theory that approaches a
conformal fixed point in the UV, rather than QCD. One may think that
this is not an important difference for the IR physics, but this is
not quite right. For example as we have seen it introduces a ``radion problem."

Thus we set out to find a potential for the dilaton that will
reproduce the  logarithmic running of the coupling. We assume that
similarly to string theory the gauge coupling is actually given by
$e^{b \phi(z)}$ (where $b$ is a numerical constant). We will find a
result consistent with expectations from string
theory.\footnote{Other discussions of backgrounds with
logarithmically running coupling can be found
in~\cite{type0,D3D7Kirsch}.}

 We assume that our action is:
\begin{equation}
S = \frac{1}{2\kappa^2} \int {\rm d}^5 x \sqrt{g} (-{\cal R}
-V(\phi)  +\frac{1}{2} g^{MN} \partial_{M} \phi \partial_{N} \phi),
\label{action2}
\end{equation}
where now we will try to determine $V(\phi)$ such that we reproduce
asymptotic freedom.  If we require that the coupling runs
logarithmically, and as usual identify the energy scale with the
inverse of the AdS coordinate $z$ we need to have a solution of the
form
\begin{equation}
e^{b \phi(z)} = \frac{1}{\log \frac{z_0}{z}},
\end{equation}
where $z_0 = \Lambda_{QCD}^{-1}$ and we do not fix $b$ {\em a
priori}.  Then going to the $y$ coordinates as usual via the
definition $e^{y/R} = z/R$, we will find
\begin{equation}
e^{b \phi(y)} = \frac{R}{y_0-y}.
\end{equation}
If we now assume that this solution follows from a superpotential
$W$ then $\phi'(y) = 6\frac{\partial W}{\partial \phi} =
\frac{1}{b(y_0-y)} = \frac{1}{bR} e^{b\phi}$. This implies that
$W(\phi) = \frac{1}{6Rb^2}e^{b \phi} + W_0$. Now that we have found
the form of the superpotential, we can easily solve for the warp
factor: $A'(y) = W(\phi(y)) = \frac{1}{6b^2(y-y_0)}+W_0$, and hence
$A(y) = A_0 + W_0 y + \frac{1}{6b^2}\log{\frac{R}{y_0-y}}$. In $z$
coordinates, this becomes:
\begin{equation}
A(z) = A_0 + W_0 R \log{z/R} - \frac{1}{6 b^2}\log{\log{z_0/z}},
\end{equation}
and hence $e^{-2A(z)} = e^{-2A_0} (R/z)^{2 W_0 R}
(\log{z_0/z})^{1/(3 b^2)}$.  From this we conclude that we should
take $A_0 = 0$ and $W_0 = \frac{1}{R}$ to get a solution that looks
AdS-like up to some powers of $\log{z_0/z}$.

The potential corresponding to this superpotential is then given by
\begin{equation}
V(\phi) =18 \left(\frac{\partial W}{\partial \phi}\right)^2 - 12 W^2
=  -\frac{1}{3 b^2 R^2} \left(
\left(\frac{1}{b^2}-\frac{3}{2}\right) e^{2 b \phi} + 12  e^{b \phi}
+ 36 b^2 \right).
\end{equation}
This is {\em particularly} simple in the case that $b = \pm
\sqrt{\frac{2}{3}}$.  In that case we have simply
\begin{eqnarray}
&&V(\phi) =   -\frac{6}{R^2} e^{\pm \sqrt{\frac{2}{3}}\phi} -
\frac{12}{R^2} \label{potential} \\
&&W(\phi )= \frac{1}{R} \left( \frac{1}{4} e^{\pm \sqrt{\frac{2}{3}} \phi} +1\right) \\
&&\phi = \mp\sqrt{\frac{3}{2}} \log \frac{y_0-y}{R}= \mp
\sqrt{\frac{3}{2}} \log \log \frac{z_0}{z} \label{sol1} \\ &&A
=\frac{y}{R}+\frac{1}{4}\log \frac{R}{y_0-y}=\log
\frac{z}{R}-\frac{1}{4}\log \log \frac{z_0}{z}, \label{sol2}
\end{eqnarray} and thus the metric will be
\begin{equation} ds^2 =
\left( \frac{R}{z}\right)^2 \left(\sqrt{\log{\frac{z_0}{z}}}  dx^\mu
dx^\nu \eta_{\mu\nu} -dz^2\right)= \frac{e^{-2
\frac{y}{R}}}{\sqrt{y_0-y}}dx^\mu dx^\nu \eta_{\mu\nu} -dy^2.
\end{equation}

Our dilaton in (\ref{action2}) is normalized in an unusual way,
nevertheless (\ref{potential}) is recognizable as a potential that
commonly occurs in nonsupersymmetric string theory backgrounds: a
cosmological constant plus a term exponential in the dilaton. This
is quite reasonable from the string theory perspective, where such a
term can arise from dilaton tadpoles in critical backgrounds or from
the central charge in noncritical backgrounds. In fact, the factor
$\sqrt{2/3}$ arises from string theory considerations in a simple
way, which may be an amusing coincidence or may have more
significance. Suppose that there is a noncritical string theory in 5
dimensions. Its action in string frame has the form
\cite{Polchinski}
\begin{equation}
S = \frac{1}{2\kappa_0^2} \int d^5 x (-G)^{1/2} e^{-2 \Phi} \left(C
+ R +  4 \partial_{\mu} \Phi \partial^{\mu} \Phi + \cdots \right),
\end{equation}
where $C$ is proportional to the central charge and is nonvanishing
since we are  dealing with a noncritical string. Now we go to
Einstein frame:
\begin{equation}
S = \frac{1}{2\kappa^2} \int d^5 x (-\tilde{G})^{1/2} \left(C e^{4
\tilde{\Phi}/3}  + \tilde{R} - \frac{4}{3} \partial_{\mu}
\tilde{\Phi} \tilde{\partial}^{\mu} \tilde{\Phi} + \cdots \right),
\end{equation}
and finally we note that comparing to our normalization above,
$\tilde{\Phi} =  \sqrt{3/8} \phi$, so that $e^{4 \tilde{\Phi}/3} =
e^{\sqrt{2/3} \phi}$.

\subsection{The Glueball Spectrum}

To calculate the glueball spectrum we can apply Eq.~(\ref{eq:radion})
for the background in (\ref{sol1}-\ref{sol2}). This equation (transformed to z coordinates)
reduces to (in units of $R$)
\begin{equation} \label{eq:gb}
z^2 F''(z) -z \left( 1+\frac{5}{2 \log \frac{z_0}{z}}\right) F'(z)+\left( \frac{4}{\log
\frac{z_0}{z}}+\frac{m^2 z^2}{\sqrt{\log \frac{z_0}{z}}} \right)F(z)=0.
\end{equation}
Using the shooting method again
we find (in units of $z_0^{-1}$) glueballs at 2.52, 5.45, 8.16, and
10.81. In particular this background seems to have a light mode from
the radion, but not any zero mode.

Lattice estimates put the first $0^{++}$ glueball in pure SU(3) gauge theory
at approximately 1730 MeV, and the second at about 2670 MeV,
with uncertainties of order 100 MeV \cite{Lattice}. Thus they put the ratio
of the first and second scalar glueball masses about about 1.54, whereas
we find a significantly larger value of 2.16. While the lattice errors are still
fairly large, this probably indicates that we are not so successful at precisely
determining properties of the second scalar glueball resonance. Since we
undoubtedly fail to properly describe highly excited resonances, this is
not so surprising. If we set $z_0^{-1}$ to match the lattice estimate for
the first glueball mass, we find
\begin{equation} \label{eq:setscale}
z_0^{-1} \approx 680~{\rm MeV}.
\end{equation}

One can also calculate the spectrum of spin $2^{++}$ glueball masses by
solving the fluctuations of the Einstein equation around the
background. The resulting differential equation we find is
\begin{equation}
z \log \frac{z_0}{z} f''(z)-(1+3 \log \frac{z_0}{z}) f'(z) +m^2
z\sqrt{\log\frac{z_0}{z}} f(z)=0.
\end{equation}
Using the shooting method (and imposing Dirichlet BC on the UV
boundary) we find the lightest modes at 4.03, 6.56, ... in units of
$1/z_0$. It is important to point out that the lowest spin $2^{++}$
glueball is naturally heavier in this setup than the spin $0^{++}$
glueball due to the mixing of the radion with the dilaton. In the
usual supergravity solutions the spin $0^{++}$ and $2^{++}$ glueballs usually
end up degenenerate (in contradiction to lattice simulations). This
is for example the case in the AdS black hole solution of Witten
analyzed in~\cite{SugraBH} (the additional light scalar modes
identified in~\cite{BMT} do not correspond to QCD modes).

\subsection{Power Corrections and gluon condensate}

We would now like to evaluate the gluon condensate in this theory assuming that
the the IR scale $z_0$ is fixed by the value of the lightest glueball mass.
As usual one needs to calculate the 5D action corresponding to a fixed source term turned
on for the QCD coupling and to get the condensate (one -point function) we need to differentiate
the 5D action with respect to the source.
Ordinarily, the computation  of the 5D action on a given solution in
AdS/QCD reduces to simply a boundary term. However, in our case it
is not so simple: our potential is not just a mass term, so that the bulk piece $V
- \frac{\phi}{2} \frac{\partial V}{\partial \phi}$ is not set to zero by the
equations of motion. As a result, the action also has a ``bulk" piece.

We evaluate the 5D action as  a function of $z_0$, imposing a UV
cutoff at $\epsilon$.
The action is given by:
\begin{equation}
\frac{1}{2\kappa^2} \int_{z=\epsilon}^{z_0} \left(
\frac{R}{z}\right)^5  \log \frac{z_0}{z} \left( -{\cal
R}-\frac{1}{2} z^2 \phi'(z)^2+\frac{12}{R^2} +\frac{6}{R^2}
e^{\sqrt{\frac{2}{3}}\phi (z)} \right) dz.
\end{equation}
This integral can be performed explicitly:
\begin{equation}
\frac{1}{2\kappa^2} \left[ \frac{1}{2 z^4}  +\frac{2 \log
\frac{z_0}{z}}{z^4}\right]_{z=\epsilon}^{z_0}
\end{equation}
We drop the UV divergent terms (the $1/\epsilon^4$ pieces) assuming that there will be
counter terms absorbing these. Then the explicit expression for the
action will be
\begin{equation}
S(z_0)=\frac{1}{4 \kappa^2 z_0^4}.
\end{equation}

Note that this is more easily calculated as
\begin{equation} \label{eq:actionisw}
S(z_0) = \frac{1}{2 \kappa^2} \int d^4 x~2 \sqrt{g_{4D}}~W(\phi),
 \end{equation}
 evaluated at the boundary $z = z_0$, where $g_{4D}$ is the induced 4D metric at the boundary.
(This observation has been made before in Ref.~\cite{GoWithRG}.)
To see this, we use the following relations:
\begin{eqnarray}
-{\cal R}(y) = -20 A'(y)^2 + 8 A''(y) & = & - 20 W(\phi)^2 + 48 \left(\frac{\partial W}{\partial \phi}\right)^2 \\
-\frac{1}{2} \phi'(y)^2 & = & - 18 \left(\frac{\partial W}{\partial \phi}\right)^2 \\
-V(\phi) & = & - 18 \left(\frac{\partial W}{\partial \phi}\right)^2 + 12 W^2
\end{eqnarray}
to see that $\sqrt{g} S_{5D}$, evaluated on the solution, is
\begin{equation}
e^{-4 A(y)} \left(-8 W^2 + 12 \left(\frac{\partial W}{\partial \phi}\right)^2\right) = 2 \frac{d}{dy} \left(e^{-4 A(y)} W(\phi(y))\right).
\end{equation}
This makes it clear that we can use the superpotential as a counterterm on the
UV boundary to cancel the terms diverging as $\epsilon \rightarrow 0$ (which
we dropped above.)

It turns out that $\frac{1}{\kappa^2}$ is almost precisely as in
the background without asymptotic freedom (because, for the
fluctuating modes, corrections to the wavefunction near $z = 0$ are
small $\alpha_s$ corrections), so one can still use (\ref{kap2}) to find
the value of $R^3/\kappa^2$. However, there is a slight subtlety:
we found a value for $\frac{1}{\kappa^2}$ assuming a source coupled
to ${\rm Tr} G^2$. In fact in our case we have fixed the numerical factor
in the correspondence of $e^{\sqrt{2/3} \phi}$ based on its
asymptotic behavior. Using the expression for the
coupling in a pure YM theory
\begin{equation}
\alpha_{YM}(Q)= \frac{2 \pi}{\frac{11}{3} N_c \log \frac{Q}{\Lambda_{QCD}}}
\end{equation}
and identifying $\Lambda_{QCD}=\frac{1}{z_0}$ and $Q = \frac{1}{z}$,
we have at the cutoff $z = \epsilon = \frac{1}{\Lambda}$:
\begin{equation}
e^{\sqrt{\frac{2}{3}} \phi(\epsilon)} = \frac{11 N_c}{6 \pi} \alpha_{YM}(\Lambda)
= \frac{11 N_c}{24 \pi^2} g^2_{YM}(\Lambda).
\end{equation}
Now, for a fluctuation $\varphi(z)$, we have
\begin{equation} \label{eq:phifluctuation}
e^{\sqrt{2/3} (\phi(z) + \varphi(z) e^{i q\cdot x})}
\approx e^{\sqrt{2/3} \phi(z)} (1 + \sqrt{2/3} \varphi(z) e^{i q \cdot x}).
\end{equation}
Now, $\varphi(z)$ near
$z = 0$ behaves like any massless scalar fluctuation on an AdS background,
and thus we have that it shifts the action by an amount
\begin{equation}
S_{5D} = \frac{1}{2 \kappa^2} \varphi(\epsilon)^2 \frac{-1}{32} q^4 \log q^2/\mu^2 + \cdots.
\end{equation}
The key now is to understand precisely which field theory correlator corresponds
to taking the second derivative of this expression with respect to $\varphi(\epsilon)$.
The field theory action is $-\frac{1}{4 g^2_{YM}} F^2$; effectively, we are adding a source
by taking the coefficient to be instead
$-\frac{1}{4 g^2_{YM}} (1 + \delta e^{i q \cdot x})$. Comparing this to
Eq.~\ref{eq:phifluctuation}, we see that $\delta = - \sqrt{\frac{2}{3}} \varphi(\epsilon)$.
Now, we have the two-point correlator for $G^2 = g_{YM}^{-2} F^2$:
\begin{equation}
\int d^4x \langle \frac{1}{4} G^2 (x) \frac{1}{4} G^2 (0) \rangle e^{i q x} = -\frac{(N_c^2-1)}{64 \pi^2} q^4 \log \frac{q^2}{\mu^2} +\ldots,
\end{equation}
which should correspond to taking a second derivative with respect to $\delta$ of our above result,
and so we find:
\begin{equation}
\frac{R^3}{\kappa^2} = \frac{64}{3} \frac{N_c^2 - 1}{64 \pi^2} = \frac{(N_c^2 - 1)}{3 \pi^2}.
\end{equation}
In particular, for $N_c = 3$, this means that
\begin{equation}
S_{5D} = \frac{2}{3 \pi^2 z_0^4}.
\end{equation}

In order to find the actual condensate, we have to
differentiate the action with respect to the value of
the source on the boundary. In our case the source is just
the QCD coupling itself $g_{YM}^{-2}$. Using the expression for the
coupling in a pure YM theory
\begin{equation}
\alpha_{YM}(Q)= \frac{2 \pi}{\frac{11}{3} N_c \log \frac{Q}{\Lambda_{QCD}}}
\end{equation}
and identifying $\Lambda_{QCD}=\frac{1}{z_0}$ we find that the
derivative with respect to $g_{YM}^{-2}$  (viewing $g_{YM}$ as a function of $z_0$) is
the same as $\frac{24 \pi^2}{11 N_c} z_0 \frac{d}{dz_0}$.

Putting all this together, we find
that:
\begin{equation}
\left<\frac{1}{4}  {\rm Tr}F^2\right> = \frac{(N_c^2 -1)}{12
\pi^2} \frac{24 \pi^2}{11 N_c} 4 z_0^{-4} \approx (1.19
z_0^{-1})^4.
\end{equation}
Using our estimate of $z_0$ from the glueball mass in
Eq.~\ref{eq:setscale}, we obtain:
\begin{equation}
\left<\frac{1}{4 \pi^2}  {\rm Tr}F^2\right> = \frac{1}{\pi^2}(1.19
\times 680~{\rm MeV})^4 \approx 0.043~{\rm GeV}^4.
\end{equation}
For comparison, an SU(3) lattice calculation found
$\left<\frac{\alpha_s}{\pi} G^2\right> \approx 0.10~{\rm GeV}^4$
~\cite{LatticeCondensate}.
Our result is of the same order but slightly smaller. Most
phenomenological estimates are smaller, beginning
with the SVZ result of
$\left<\frac{\alpha_s}{\pi} G^2\right> \approx 0.012~{\rm GeV}^4$,
but for pure Yang-Mills the value is expected to increase~\cite{SVZ}.

\subsection{Relation to Analytic Perturbation Theory}

The QCD perturbation series is an asymptotic expansion of some
unknown function, and  the divergence at $\Lambda_{QCD}$ signals
only a breakdown of perturbation theory, not a meaningful infinity.
In particular, it has been proposed that the pole of the logarithm
be cancelled by additional terms to produce an ``analytic
perturbation theory." See, for instance, the work of Shirkov and
Solovtsov \cite{ShirkovSolovtsov} and related literature (of which
there is too much to give an exhaustive account here). It is
interesting that our holographic equations produce a result along
these lines, when interpreted in a particular way.

To see this, note that  our identification of the coordinate $z$
with the inverse of a renormalization group scale $\mu$ is only
clearly defined in the far UV (near $z = 0$). In fact, when we take
as a metric ansatz ${\rm d}s^2 = \exp(-2 A(y)) {\rm d}x^2 + {\rm
d}y^2$, it is more reasonable to interpret $A(y)$ as $-\log \mu R$, so that
the 4D part of the metric goes like $\mu^2 {\rm d}x^2$. Our
expression for the QCD coupling in terms of the dilaton $\exp{\sqrt{2/3} \phi(y)}$
is given by $\frac{R}{y_0 - y}$, which blows up at a finite coordinate in exactly the way that the
one-loop QCD beta function tells us $\alpha_s$ should blow up at $\mu =\Lambda_{QCD}$.
On the other hand, using the modified identification of the energy scale
explained above,
\[ A(y) \leftrightarrow -\log \mu R, \]
that is by identifying {\em the warp factor} (instead of $y$)
as the logarithm of the energy scale,
we find that $y \rightarrow y_0$ corresponds to $\mu \rightarrow 0$.
Thus we can view $\exp{\sqrt{2/3} \phi(y)}$ as providing a formula
for $\alpha_s(\mu)$ that is smoothly defined at all $\mu$, which
blows up as a power law in the deep infrared $\mu \to 0$ (instead of
$\mu \to \Lambda_{QCD}$) and reduces to the
perturbative result at large $\mu$.

In particular, one can  solve for $\alpha_s(\mu)$ according to this
prescription. The relevant expressions
\begin{eqnarray}
&& \frac{6\pi}{11 N_c} \alpha_s^{-1} (\mu) =(y_0-y(\mu))/R \\
&& -\log \mu R = \frac{y(\mu )}{R}+\frac{1}{4} \log \frac{R}{y_0-y(\mu )} \\
&& \Lambda_{QCD}=z_0^{-1}=\frac{1}{R} e^{-y_0/R} \end{eqnarray}
can be inverted to find
\begin{equation}
\frac{1}{\alpha_s(\mu)} = \frac{11 N_c}{24\pi} W(4 \mu^4 /
\Lambda_{QCD}^4),
\end{equation}
where $W(y)$ is the Lambert W-function
\cite{LambertW}, that is, the principal value of the solution to $y
= x \exp(x)$. In fact, the Lambert W-function appears similarly in
the analytic perturbation theory approach \cite{APTLambert},
although the form of $\alpha_s(\mu)$ is slightly different there.
Nonetheless, our results are suggestive of a role for the
backreaction on the metric as enforcing good analytic properties,
which deserves further attention.

\section{Effects of the Tr($F^3$) condensate}
\setcounter{equation}{0}
\setcounter{footnote}{0}

In pure Yang-Mills, there is an operator of dimension 6,  ${\cal
O}_6 = f^{abc}F^{a}_{\mu \nu}F^b_{\nu \rho}F^c_{\rho\mu}$. This
operator will get a condensate which modifies the OPE at higher
orders. We wish to investigate the size of the corrections on the
glueball masses in our approach, to understand how stable the
numerics are. Thus we add a new field $\chi$, and we wish to modify
the superpotential. We should still have in the potential terms
$\Lambda$ and $C \exp(b \phi)$, as before. However, now we should
also have a mass term for $\chi$, with $m_{\chi}^2 = 6(6-4) = 12$ by
the AdS/CFT correspondence. On the other hand, our theory is not
conformal, and ${\cal O}_6$ has an anomalous dimension proportional
to $\alpha_s$, suggesting we should also have terms in the potential
that couple $\phi$ and $\chi$.

For now, we will make no attempt to constrain all the higher-order
terms in the 5D action coupling $\phi$ and $\chi$. Instead, we seek
a superpotential with the properties discussed above, as a first
approximation. Luckily, there is a superpotential which allows the
profiles of the scalars and the warp factor to be found
analytically. Our motivation for this particular choice is that it
allows an analytic solution and has the desired properties. On the
other hand, it resembles certain superpotentials that arise in
gauged supergravity~\cite{Massimo}:
\begin{equation}
W(\phi, \chi) = \frac{1}{4} \exp\left(\sqrt{\frac{2}{3}} \phi \right) + \cosh (B \chi).
\end{equation}
(We will see shortly that $B = 1$.) The corresponding potential is then:
\begin{eqnarray}
V(\phi, \chi) & = & 18 \left[ \left(\frac{\partial W}{\partial \phi}\right)^2 + \left(\frac{\partial W}{\partial \chi}\right)^2 \right] - 12 W^2 \nonumber \\
&  =  & - 12 - 6 e^{\sqrt{\frac{2}{3}}\phi} + (18 B^4 - 12 B^2) \chi^2 - 3 B^2 e^{\sqrt{\frac{2}{3}}\phi} \chi^2 + {\cal O}(\chi^4).
\end{eqnarray}
We find that $\phi(y)$ is as before, whereas $\chi(y)$ is given by $\chi'(y) = 6 \frac{\partial W}{\partial \chi} = 6 B \sinh(B \chi)$. But this is nearly identical to the equation we solved to find $\phi(y)$ in the case {\em without} log running. In particular, this means that
\begin{equation}
\chi(y) = \log\left(\tanh \frac{3(y_1 - y)}{R}\right),
\end{equation}
where we have chosen $B = 1$ to ensure that at small $z$, $\chi(z)
\sim z^6$.  In fact we can check this, as in the potential above we
expect $(18 B^4 - 12 B^2) \chi^2 = 6 \chi^2$, confirming that we
want $B = 1$.

Finally, we evaluate the warp factor,  using $A'(y) = W(\phi(y),\chi(y))$:
\begin{equation}
A(y) = -\frac{1}{6} \log\left(\cosh \frac{3(y_1-y)}{R} \sinh \frac{3(y_1-y)}{R}\right) + \frac{1}{4} \log\left(\frac{R}{y_0-y}\right) + \frac{y_1}{R} - \frac{\log 2}{3},
\end{equation}
where the first term replaces the $y$ of our solution without the
inclusion of ${\cal O}_6$,  but deviates from it in the infrared.
(The constant terms correct for a constant difference between $y$
and the first term, in the far UV.) The solution in the $z$
coordinates is given by
\begin{eqnarray}
&& ds^2= \left( \frac{R}{z}\right)^2 \left[ \left(\log
\frac{z_0}{z}\right)^{\frac{1}{2}}
\left(1-\frac{z^{12}}{z_1^{12}}\right)^{\frac{1}{3}} dx^\mu dx^\nu
\eta_{\mu\nu} -dz^2 \right] \\
&& \phi (z)=-\sqrt{\frac{3}{2}} \log \log \frac{z_0}{z} \\
&& \chi (z)= \log \frac{1-\frac{z^6}{z_1^6}}{1+\frac{z^6}{z_1^6}}.
\end{eqnarray}

At this point the reader should be notice certain issues in our
calculation. First, while it  is true that the potential $V(\phi,
\chi)$ couples $\phi$ and $\chi$ and includes a term suggestive of
an anomalous dimension, the solutions for $\phi(y)$ and $\chi(y)$
themselves are completely decoupled! This, however, is not really a
concern: the backreaction on the metric feels all of the terms in
the potential. In other words, it is {\em only through A(y)} that
the anomalous dimension is manifest. If, as suggested earlier, we
interpret $A(y)$ as $-\log \mu$, then $\phi(\mu)$ and $\chi(\mu)$
will feel the effect of the anomalous dimension.

Another issue is that we have sacrificed precise agreement with
perturbation theory for the  sake of having a simple, solvable
example. We have arranged to get the logarithmic running of
$\alpha_s$, the proper scaling dimension of ${\cal O}_6$, and an
anomalous dimension term for ${\cal O}_6$ which is proportional to
$\alpha_s$. On the other hand, we have not been careful to match the
coefficient in this anomalous dimension. Details of the
OPE and anomalous dimension
for this operator can be found in Ref.~\cite{ThreeGlue};
eventually one would want a model that matches them.
Of course, in our discussion of
$\alpha_s(\mu)$ earlier, we also had a second $\beta$ function
coefficient that did not match. This suggests that our analytically
solvable superpotentials, while useful for a preliminary study,
probably need to be replaced by a more detailed numerical study
based on a more careful matching of the holographic renormalization
group. Nonetheless, the disagreement appears only at higher orders
of perturbation theory, and we can already use our preliminary
superpotential $W(\phi,\chi)$ to get some sense of the stability of
spectra calculated in holographic models.

\subsection{Gubser's Criterion: Constraining $z_1/z_0$} \label{sec:Gubser}

In our solution, $\phi(z)$ blows up at $z = z_0$ while $\chi(z)$
blows up at $z = z_1$. The space will shut off at whichever
of these is encountered first. Intuitively it is clear that if
the dimension 6 condensate is to make a relatively small
correction to the results we have already obtained using
only the dimension 4 condensate, we should have $z_1 > z_0$,
so that $\chi(z)$ remains finite over the interval where
the solution is defined.

In fact there is a conjecture that will enforce this condition.
Namely, Gubser in Ref.~\cite{GubserSingularities} has
proposed that curvature singularities of the type arising
in the geometries we are considering are allowed only
if the scalar potential is bounded above when evaluated
on the solution. By ``allowed", one should understand that
in these cases one expects the singularity to be resolved
in the full string theory; geometries violating the criterion
are somehow pathological. The conjecture is based on
some nontrivial consistency checks involving
considerations of finite temperature and examples
from the Coulomb branch in AdS/CFT, but it is not proven.
In any case we will assume for now that it holds for any
geometry that can be properly thought of as dual to
a field theory. It is clear that our original solution
involving only $\phi$ satisfies the criterion: in that case we
had $V(\phi) = - 12 - 6 e^{\sqrt{2/3} \phi} < 0$.

The case with $\chi$ is more subtle. We have
\begin{eqnarray}
V(\phi(z),\chi(z)) & = & -15 - 6 e^{\sqrt{\frac{2}{3}} \phi(z)} \cosh \chi(z) + \cosh (2 \chi(z)) \nonumber \\
& = & -6~\frac{\left(1 - \left(\frac{z}{z_1}\right)^{24}\right)}{\left(1-\left(\frac{z}{z_1}\right)^{12}\right)^2 \log\frac{z_0}{z}} - 12~\frac{1 - 4 \left(\frac{z}{z_1}\right)^{12} + \left(\frac{z}{z_1}\right)^{24}}{\left(1-\left(\frac{z}{z_1}\right)^{12}\right)^2 }
\end{eqnarray}
Clearly as $z_1 \rightarrow \infty$ we recover the previous solution
and Gubser's criterion is satisfied. On the other hand, as soon as $z_1 < z_0$,
$V$ begins to attain large positive values as $z \rightarrow z_1$.
The reason is simple: the function $1 - 4 x^{12} + x^{24}$ has a zero at $x \approx 0.9$,
so the second term above can attain positive values when $z \approx z_1$,
and the denominator will attain arbitrarily small values provided the singularity
at $z_1$ is reached (i.e. $z_1 < z_0$). In fact large positive values of $V$
are attained if and only if $z_1 < z_0$.

In short, Gubser's criterion limits us to precisely those solutions which can
be viewed in some sense as a perturbation of our existing solution.

\subsection{Condensates}

To calculate the condensate we need to again evaluate the classical
action for the solution, which in our case is given by
\begin{eqnarray}
S= && \frac{1}{2\kappa^2} \int_{\epsilon}^{z_0} \left(
\frac{R}{z}\right)^6 \log \frac{z_0}{z} \left(
1-\frac{z^{12}}{z_1^{12}}\right)^{\frac{2}{3}}\left[ -\frac{1}{2}
(\phi'^2+\chi'^2) \left( \frac{z}{R}\right)^2 \right. \nonumber \\
&& \left. +12 +6 e^{\sqrt{\frac{2}{3}} \phi} \cosh \chi +6 \sinh^2
\chi -{\cal R}\right].
\end{eqnarray}

Again dropping the UV divergent terms we find either using (\ref{eq:actionisw}) or by
direct integration
\begin{equation}
S=\frac{1}{2\kappa^2} \frac{(z_1^{12}-z_0^{12})^{\frac{2}{3}}}{2
z_0^4 z_1^8}
\end{equation}
Following the steps for calculating the condensate for the single
field case we find that the modified condensate is given by
\begin{equation}
\left<\frac{1}{4}  {\rm Tr}F^2\right> = \frac{(N_c^2 -1)}{3 \pi^2}
\frac{24 \pi^2}{11 N_c} z_0 \frac{d}{dz_0} \left[ \frac{(z_1^{12}-z_0^{12})^{\frac{2}{3}}}{4 z_0^4 z_1^8} \right] \approx (1.19 z_0^{-1})^4 \frac{1+\left( \frac{z_0}{z_1}\right)^{12}}{\left[1-\left( \frac{z_0}{z_1}\right)^{12}\right]^{\frac{1}{3}}}.
\end{equation}

One can see that for $z_1>z_0$ (as expected from the criterion of
Sec.~\ref{sec:Gubser}) this condensate is very insensitive to the actual value of
$z_1$. In order to actually fix $z_1$ one would have to calculate the
second condensate $\langle {\rm Tr} F^3\rangle $ and compare it to the
lattice results. However, there are no reliable lattice estimates for this
condensate available.

It is also plausible that if we account more properly for the anomalous
dimension of ${\rm Tr} F^3$ and for matching of perturbative corrections
to the OPE, we will be able to select a solution without this ambiguity.
However, constructing such a solution appears to require a numerical
study, which we will leave for future work.

\subsection{Glueball spectra}

Now that we have the background deformed by condensates of Tr $F^2$
and Tr $F^3$, we can  again compute the glueball spectrum. What we would like
to check is how sensitive the results are to the value of $z_1$.
To simplify the numerical problem we are assuming that the low-lying
glueball modes are still predominantly contained in the $\phi$ and $A$ fields,
and that the leading effect of turning on the $\chi$ field is to modify the
gravitational background $A(y)$. Using this approximation we find the following
equation satisfied by the glueball wave functions (again in $z$ coordinates and units
of $R$):
\begin{eqnarray}
&& z^2 F''(z) -z \left( 1+\frac{5}{2 \log \frac{z_0}{z}}-\frac{4}{1-\frac{z_1^{12}}{z^12}}
\right) F'(z)
+\nonumber \\ && \left( \frac{4}{\log \frac{z_0}{z}}
\frac{1+\frac{z^{12}}{z_1^{12}}}{1-\frac{z^{12}}{z_1^{12}}}
+\frac{m^2 z^2}{\left(1-\frac{z^{12}}{z_1^{12}}\right)^{\frac{1}{3}} \sqrt{\log \frac{z_0}{z}}}
-\frac{96 z^{12}}{z_1^{12} \left(1-\frac{z^{12}}{z_1^{12}}\right)^2 }
\right)F(z)=0.
\end{eqnarray}

One can see that the equation reduces to (\ref{eq:gb}) in the limit when $z_1 \gg z_0 \geq z$.
By again numerically solving this equation for various values of $z_1/z_0 >1$ we find that
the glueball eigenvalues are very insensitive to the actual value of $z_1$ as long as
$z_1$ is not extremely close to $z_0$. For example,  the lightest eigenvalue at $2.52/z_0$ increases
by less than a percent while lowering $z_1/z_0$ from $\infty$ to $1.5$. For $z_1/z_0 =1.1$
the lightest mass grows by 3 percent, while for the extreme value of $z_1/z_0 =1.01$ the growth is
still just 9 percent. The glueball mass ratios are even less sensitive to $z_1$: the ratio
of the first excited state to the lightest modes decreases by about 3 percent when changing
$z_1/z_0$ from $\infty$ to $1.01$. Thus we conclude that the predictions for glueball spectra
presented in the previous section are quite robust against corrections from higher condensates.

\section{Linearly confining backgrounds?}
\setcounter{equation}{0} \setcounter{footnote}{0}

One of the most problematic aspects of holographic QCD is the deep
IR physics: one  expects Regge behavior from states of high angular
momentum, and a linear confining potential. The solutions presented
in this paper show many qualitative and quantitative similarities
with ordinary QCD. However, they do not properly describe the highly
excited glueball states. Since there is a singularity at a finite
distance, the characteristic mass relation for highly excited
glueballs will be that of ordinary KK theories (in this respect the
theory is similar to the models with an IR brane put in by hand)
$m_n^2 \sim n^2/z_0^2$, instead of the expected Regge-type behavior
$m_n^2 \sim n/z_0^2$~\cite{Shifman05}. Experimental and lattice data
suggest that linear confinement effects persist further into the UV
than one might expect, and already the light resonances observed in
QCD seem to fall on Regge trajectories. Regge physics arises
naturally from strings; in our approach, the more massive
excitations of the 5D string correspond to high-dimension operators
on the field theory side. To describe linear confinement and Regge
physics accurately, then, it is conceivable that one must take into
account the effects of a large number of operators. Thus we are led
to seek alternative, but still well-motivated, approaches to the
deep IR physics. One approach is simply to demand that the 5D fields
have IR profiles that provide the desired behavior, as in
Ref.\cite{KarchRegge}. However, one would like to have a dynamical
model of this effect.  Here we first check that the backgrounds used
in sections 4-5 do not reproduce such IR profiles. However
in~\cite{KarchRegge} it was suggested suggested that tachyon
condensation might provide the appropriate dynamics. We provide a
simple modification of our model possibly substantiating this idea,
and speculate on its relation to known gauge theory effects: namely,
UV renormalons and other $1/Q^2$ corrections as discussed by
Zakharov and others~\cite{QSquaredCorrections}.

\subsection{No linear confinement in the dilaton-graviton system}

We first try to find other solutions that would not have a
singularity at a finite distance, even for the action
(\ref{action2}) since until now we have found only a particular
solution for (\ref{action2}) using a superpotential.  However in
general there should be an infinite family of superpotentials giving
the same potential. For the general case we cannot find the other
solutions analytically for all values of the coordinates. We can,
however, attempt to solve the equations of motion analytically close
to the UV boundary. A similar approach has been taken in
ref.\cite{Grena} for a type 0 string theory containing a tachyon,
considered to be dual to non-supersymmetric SU(N) Yang-Mills with
six adjoint scalars. In our case we will not consider a tachyon for
now, as we consider 5D fields to be in one-to-one correspondence
with gauge invariant operators in the 4D dual. There are some
orientifold theories in type 0 that have no bulk tachyon, so our
approach is not {\em a priori} meaningless.

In order to avoid the change of variables needed to achieve an
asymptotically  AdS metric (plus power corrections), we solve the
equations beginning from the explicitly conformally flat
parametrization of the metric,
\begin{equation}
{\rm d}s^2 = e^{-2 A(z)} \left(\eta_{\mu \nu}{\rm d}x^\mu {\rm
d}x^\nu - {\rm d}z^2\right).
\end{equation}
Then for the Einstein equations and the equation of motion for
$\phi$ we find  (using $'$ to denote derivatives with respect to
$z$):
\begin{eqnarray}
\phi''(z) -3 A'(z) \phi'(z) - e^{-2A(z)} \partial V(\phi)/\partial \phi = 0, \\
6 A'(z)^2 - (1/2) \phi'(z)^2 + e^{-2A(z)} V(\phi) = 0, \\
3 A'(z)^2 -3 A''(z) + (1/2) \phi'(z)^2 + e^{-2 A(z)} V(\phi) = 0.
\end{eqnarray}
We again use the same potential $V(\phi) = -6 e^{\sqrt{2/3} \phi(z)}
- 6$ (but no longer assume the simple expression for the
superpotential), and solve the equations in the UV to obtain
\begin{eqnarray}
\phi(z) = \sqrt{\frac{3}{2}} \left(-\log \log\frac{1}{z} +  \frac{3}{4}\frac{\log\log\frac{1}{z}}{\log\frac{1}{z}}  + \frac{\gamma}{\log\frac{1}{z}} + \cdots \right), \\
A(z) = \log\frac{1}{z} + \frac{1}{2 \log\frac{1}{z}} - \frac{3}{8}
\frac{\log\log\frac{1}{z}}{\left(\log\frac{1}{z}\right)^2} +
\frac{\frac{\gamma}{2}+\frac{33}{48}}{\left(\log\frac{1}{z}\right)^2}
+ \cdots.
\end{eqnarray}
The omitted terms are those which are smaller as $z \rightarrow 0$,
i.e. power corrections  in $z$ or higher powers in $(-\log z)^{-1}$.
Here $\gamma$ is a parameter that is undetermined by the UV
equations of motion, just as in the solution of ref.\cite{Grena}.
Further, we find that if we add small perturbations to
$\exp(\sqrt{2/3} \phi)$ and to $A(z)$, an ${\cal O}(z^4)$ power
correction to the former is allowed while only constant corrections
to the latter are allowed. It is reassuring that we find a power
correction suitable for the gluon condensate. Ref.\cite{Grena} found
also a correction of order $z^2$ and interpreted it as a UV
renormalon\footnote{For a review of the physics of the UV
renormalon, see Ref.~\cite{UVrenormalon}.} (whereas the power
correction we have considered may be thought of as an IR
renormalon). The tachyon was crucial to the appearance of the UV
renormalon term, which might have interesting implications.

In order to find a linearly confining solution, we would need to
connect these UV solutions to solutions in the $z\to \infty$ IR
region, which would also give linear confinement. Suppose we then
search for a solution that is valid at large $z$. In the case that
there is no  cosmological constant, there is a linear dilaton
solution; this is clear since our equations are those for a
noncritical string. The linear dilaton persists in the presence of a
cosmological constant as a solution at large $z$; the equations are
satisfied up to terms exponentially small at large $z$ by a linear
dilaton on flat 5D space. Such a linear dilaton background does {\em
not} give a confining solution. Furthermore, one can check that no
other power-law growth in $z$ is allowed. In particular, the $z^2$
behavior as in Ref.~\cite{KarchRegge} is not a solution to our
equations of motion. Thus we conclude that none of the solutions of
the action (\ref{action2}) will result in a linearly confining
background.

\subsection{Linear confinement from the tachyon-dilaton-graviton system?}

Although we do not understand how to {\em systematically} approach
the study of UV renormalon  corrections, it is at least
superficially plausible that they are associated with a closed
string tachyon. This is a natural scenario to consider, since the UV
renormalon is associated with $Q^{-2}$ corrections.
Ref.~\cite{Grena} found such an association in a concrete string
theory model. The study of such corrections has been discussed in
great detail in the QCD literature, which links the idea of the UV
renormalon to the quadratic corrections associated with the QCD
string tension and with a hypothetical nonperturbative tachyonic
gluon mass, as well as with monopoles and vortices; we can only
refer to a small sample of that
literature~\cite{QSquaredCorrections}. Holographically, effects
associated with dimension 2 corrections would appear to be
associated with tachyonic physics. Interestingly, it has been
proposed that the UV renormalon is associated with a nonvanishing
value of $\left<A^2\right>_{min}$ where the minimization is over
gauge choices~\cite{ASquared}. Such a dimension 2 condensate could
plausibly be associated holographically to a closed string tachyon
(one that saturates the Breitenlohner-Freedman
bound~\cite{BreitenlohnerFreedman}), though it does contradict the
picture of holographic fields as being associated to {\em
gauge-invariant, local} operators in the field theory. The minimum
value of the $A^2$ condensate can be expressed in terms of {\em
nonlocal}, gauge-invariant operators, beginning with $F_{\mu
\nu}(D^2)^{-1} F^{\mu \nu}$~\cite{NonlocalDim2}. The nonlocality of
this operator is perhaps suggestive of the long-distance, stringy
effects of the flux tube needed to describe excited hadrons.

All of these hints of the importance of $Q^{-2}$ corrections suggest
that we take the idea of a closed string tachyon seriously, despite
the lack of a gauge-invariant local operator for it to match to.
Perhaps this indicates that certain 5D corrections can somehow be
``resummed" into the effects of a tachyonic field. The linear
confining potential of QCD has been suggested to relate to an
$\exp{cz^2}$ background in 5D~\cite{KarchRegge}. We would like to
have a dynamical solution incorporating both asymptotic freedom in
the UV and linear confinement in the IR. We have observed that the
cosmological constant does not destroy the existence of a linear
dilaton solution at large $z$ (up to small corrections). This
suggests that if we have a theory with a quadratic tachyon profile
at zero cosmological constant, such a solution may persist in the
presence of a cosmological constant.

The action of the bosonic noncritical string theory including the
leading $\alpha'$ corrections, using a sigma model approach, is
(after transforming to Einstein frame), according to
\cite{AndreevTachyon}:
\begin{eqnarray}
&& \frac{1}{2\kappa^2} \int d^5 x \sqrt{g} \left[
e^{\frac{4}{3}\Phi} \frac{4}{\alpha'} (\lambda -1) -2 {\cal R}
+\frac{8}{3} (\partial \Phi)^2 +e^{-t} \left( \frac{4}{\alpha'}
(1+t-\frac{1}{2} \lambda )e^{\frac{4}{3} \Phi} +{\cal R} \right.
\right. \nonumber \\ && \left. \left. -\frac{4}{3} (\partial \Phi)^2
-\frac{4}{3}
\partial \Phi
\partial t +(\partial t)^2\right) \right].
\end{eqnarray}
To this action we are adding a cosmological constant (which we
assume could come from a 5-form flux) and adjust the parameters such
that with the $t\to 0$ substitution we recover the action considered
in the previous section. The resulting action is:
\begin{eqnarray}
&& \frac{1}{2\kappa^2} \int d^5 x \sqrt{g} \left[
e^{\frac{4}{3}\Phi} \frac{12}{R^2} \frac{\lambda
-1}{\lambda}+\frac{12}{R^2} -2 {\cal R} +\frac{8}{3} (\partial
\Phi)^2 +e^{-t} \left( \frac{12}{R^2 \lambda} (1+t-\frac{1}{2}
\lambda )e^{\frac{4}{3} \Phi} +{\cal R} \right. \right. \nonumber \\
&& \left. \left. -\frac{4}{3} (\partial \Phi)^2 -\frac{4}{3}
\partial \Phi
\partial t +(\partial t)^2\right) \right].
\end{eqnarray}
The resulting equations of motion for the metric ansatz $ds^2= e^{2
A(z)} (dx^2-dz^2)$ are:
\begin{eqnarray}
&& \frac{4}{3} e^{2 A} e^{\frac{4}{3}\Phi}
\frac{12}{R^2\lambda}(\lambda -1)+\frac{16}{3} ( 3 A' \Phi'+\Phi'')+
e^{-t} \left( e^{\frac{4}{3}\Phi+2A} \frac{12}{R^2}
\frac{1}{\lambda} (1+t-\frac{1}{2}\lambda) \frac{4}{3} +\frac{8}{3}
t'\Phi' \right. \nonumber \\ && \left.  -8 A' \Phi' -\frac{8}{3}
\Phi'' +\frac{4}{3} t'^2-4 A't'\right)=0\\
&& \frac{12}{R^2\lambda} e^{\frac{4}{3}\Phi +2A}
(\frac{\lambda}{2}-t) -12 A'^2-8 A''-\frac{4}{3} \Phi'^2-t'^2 -4
A'\Phi' -\frac{4}{3} \Phi'' +2 t''+ 6 A' t' =0 \nonumber \\ 
\end{eqnarray}
\begin{eqnarray}
&& -6 A'^2(2-e^{-t})+\frac{6}{R^2} e^{2
A}(1+\frac{1}{\lambda}e^{\frac{4}{3}\Phi} (\lambda -1+e^{-t}
(1-\frac{\lambda}{2} +t))+\Phi'^2 (\frac{2}{3}-\frac{4}{3}
e^{-t})+\nonumber \\
&& \frac{1}{2} e^{-t}(-\frac{4}{3} \Phi' t' +t'^2)=0.
\end{eqnarray}
Here $\lambda = (26-D)/3=21/3$. One can show that for large $z$ the
leading order solution to these equations is given by
\begin{equation}
\Phi \to \Phi_0, \ \ A\to A_0, \ \ t\to -\frac{3}{\lambda} e^{2 A_0
+\frac{4}{3}\Phi_0} z^2+\frac{1}{2} (\lambda -2).
\end{equation}
In order to serve our purposes there must be a solution
interpolating between logarithmic running on AdS in the UV and the
above confining background (with flat space and constant dilaton) in
the IR. It would be interesting to numerically study these
equations, as well as similar systems in the type 0 string. The IR
solution above has the property that the $e^{-t}$ terms are {\em
growing} at large $z$, whereas the proposed background in
Ref.~\cite{KarchRegge} has an exponential {\em shutoff} of the
metric, which is not a solution of the above equations. However, the
sign does not appear to affect the existence of Regge physics.
Regardless, we intend these remarks not as a definitive statement,
but as a provocative hint that further studies of tachyon dynamics
could be useful for understanding QCD.

\section{Conclusions and Outlook}
\setcounter{equation}{0}
\setcounter{footnote}{0}

In this paper we have clarified a number of aspects of AdS/QCD, and
sketched a concrete program for  computing in the holographic model
and estimating associated errors. We have also explained that deep
IR physics, associated with high radial excitations or large angular
momentum, is troublesome in this framework. The underlying reason is
that the OPE does not reproduce the quadratic corrections associated
with this physics. We have suggested that a closed string tachyon
can reproduce much of the underlying dynamics, but at this point
that is a toy model and it is not clear how to systematically apply
the idea to computations.

Our results suggest a number of directions for new work. One obvious
task is to extend these  results to theories with flavor. This
should be straightforward, although there are potential numerical
difficulties. It would be particularly interesting to see if one can
obtain results for mixing of glueballs with $q\bar{q}$ mesons
without large associated uncertainties.

Another direction is to make more explicit the connection between
the renormalization group and  the holographic dual. The 5D action
has an interpretation as the 4D generating functional $W[J]$,
whereas some numerical attempts involving truncations of exact RGEs
have focused on computing the Legendre transform of this quantity,
$\Gamma[\phi]$. There are subtle issues of nonperturbative
gauge-invariant regulators that must be considered in such studies,
but it is conceivable that such existing work could be reinterpreted
as a computation of background profiles for various fields, about
which we could then compute the spectrum of excitations and
couplings with the usual 5D techniques. We hope to clarify this
relationship in a future paper. It is also interesting in this
context to think about how the holographic renormalization group
might relate to the ``analytic perturbation theory" framework.

Finally, the intricate story relating $1/q^2$ corrections, UV
renormalons, vortices, linear  confinement, and the closed string
tachyon is very interesting and still rather poorly understood. A
better understanding of these quantities and their relationships
could elucidate the confining dynamics of QCD, and also shed light
on closed string tachyon condensation in general (with possible
applications to cosmology).

\section*{Acknowledgments}
We thank Josh Erlich, Andreas Karch and Ami Katz for illuminating
discussions of AdS/QCD. The research of C.C. is supported in part by
the DOE OJI grant DE-FG02-01ER41206 and in part by the NSF grant
PHY-0355005. M.R. is supported by a National Science Foundation
Graduate Research Fellowship and in part by the NSF grant
PHY-0305005.

\end{document}